# Negative refraction index of the mesoscopic left-handed transmission line in the thermal Fock state

Shun-Cai Zhao · Hong-Wei Guo · Xiao-Jing Wei

**Abstract** Negative refractive index (NRI) of the quantized lossless mesoscopic left-handed transmission line (LHTL) is deduced numerically in thermal Fock state. Some specific quantum features of NRI dependent the temperature, frequency of the electromagnetic wave and photon numbers, and quantum fluctuations are shown in the lossless LHTL. The results are significant for the miniaturizing applications of LHTL and quantum circuits.

## 1 Introduction

The theoretical speculation of negative refractive index materials (NRM) proposed by V. Veselago[1] in 1968, in which several fundamental phenomena occurring in or in association with NRM were predicted, such as the negative Goos-Hänchen shift[2], amplification of evanescent waves[3], reversals of both Doppler shift and Cerenkov radiation[1], sub-wavelength focusing[4] and so on. Some typical approaches for NRM can be summarized as artificial structures such as metamaterials[5–7] and photonic crystals[8–10], chiral materials[11] and photonic resonant media[12,13]. Although very exciting from a physics point of view, the artificial structures seem to be of little practical interest for engineering applications because

Supported by the National Natural Science Foundation of China ( Grant Nos. 61205205 and 6156508508 ), the General Program of Yunnan Provincial Research Foundation of Basic Research for application, China ( Grant No. 2016FB009 ) and the Foundation for Personnel training projects of Yunnan Province, China ( Grant No. KKSY201207068 ).

S.C. Zhao
Department of Physics, Faculty of Science, Kunming University of Science and Technology, Kunming, 650500, PR China
E-mail: zhaosc@kmust.edu.cn

H. W. Guo · X. J. Wei
Department of Physics, Faculty of Science, Kunming University of Science and Technology, Kunming, 650500, PR China

of these resonant structures exhibiting high loss and narrow bandwidth consequently. Due to the weaknesses of resonant-type structures, three groups almost simultaneously in June 2002 introduced a transmission line (TL) approach of N-RM: Eleftheriades et. al[15,16], Oliner[17] and Caloz et. al.[18,19]. LHTL initially the non-resonant-type one, is perhaps one of the most representative and potential candidates due to its low loss, broad operating frequency band, as well as planar configuration[20,21], which is often related with easy fabrication for N-RI applications in a suite of novel guided-wave[22,23], radiated-wave[24,25], and refracted-wave devices and structures[26–28].

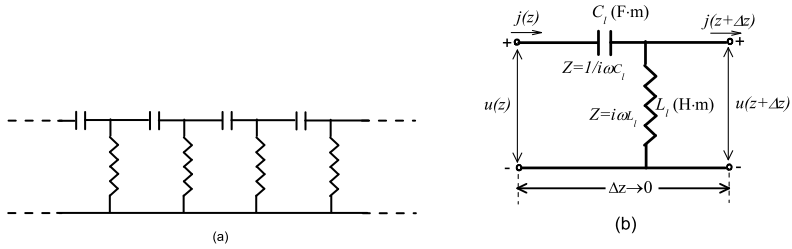

**Fig. 1** (a) Equivalent circuit model for the uniform LHTL, (b) Unit cell circuit model for a uniform LHTL.

In the meanwhile, with the rapid development of nanotechnology and nanoelectronics [29], the integrated circuits and components have minimized towards atomic-scale dimensions[30,31] in the last a few decades. When the scale of fabricated electric materials reached to a characteristic dimension, namely, Fermi wavelength, quantum mechanical properties of mesoscopic physics[32,33] become important while the application of classical mechanics fails. The miniaturizing applications would be undoubtedly a persistent trend for LHTL, and the quantum properties may cast some influence on NRI. So it's significant to investigate the quantum properties of NRI in the LHTL.

Thus, from the point of against this miniaturization challenge, this paper exploits the quantum effect on NRI in the thermal Fock state of the mesoscopic LHTL. The fundamental features of LHTL in Fig.1(a) are straightforwardly derived by elementary TL theory. It consists of lossless per-unit equivalent circuit models (Fig.1.(b)) of the series-C/shunt-L prototype associated with NRI in microwave frequency band[22,34]. The per-unit-length inductance $L_l$ (H· m) and capacitance $C_l$ (F· m) are $L_l = L'_l \cdot \Delta_z$ and $C_l = C'_l \cdot \Delta_z$, respectively. So we have the impedances $Z = 1/i\omega C_l$ ($\Omega/m$) and admittances $Y = 1/i\omega L_l$ (S/m).

According to the Fig.1(b), the complex propagation constant $\gamma$, the propagation constant $\beta$, the characteristic impedance $Z_l$, the phase velocity $v_p$, and the group velocity $v_g$ with the the equivalent constitutive permittivity and permeabil-ity of the unit cell equivalent circuit for LHTL are given by[34]

$$\gamma = -i\frac{1}{\omega\sqrt{C_l L_l}}, \beta = -\frac{1}{\omega\sqrt{C_l L_l}} < 0,$$

$$Z_l = \sqrt{\frac{L_l}{C_l}},$$

$$v_p = \frac{\omega}{\beta} = -\omega^2\sqrt{C_l L_l} < 0,$$

$$v_g = (\frac{\partial \beta}{\partial \omega})^{-1} = \omega^2\sqrt{C_l L_l} > 0, \quad (1)$$

$$\mu(\omega) = -\frac{1}{\omega^2 C_l}, \epsilon(\omega) = -\frac{1}{\omega^2 L_l}$$

According to Kirchhoff's law, the classical differential equations of motion of Fig.1(b) are

$$\frac{d^2 u(z)}{dz^2} = -\gamma^2 u(z) \quad (2)$$

$$\frac{d^2 j(z)}{dz^2} = -\gamma^2 j(z) \quad (3)$$

where $u$ and $j$ are the position-dependent voltage and currents $u = u(z)$ and $j = j(z)$ along the line, respectively. where $\gamma$ is the complex propagation constant. And the walking-wave solutions to Eq(2) and Eq(3) reach as $j(z) = A\exp(-i\gamma z) + A^*\exp(i\gamma z), u(z) = B\exp(-i\gamma z) + B^*\exp(i\gamma z)$, in which $A^*$ ($B^*$) are the conjugate complexes of A (B). We adopt the quantization method similar to Louisell [35] to achieve the current operator. In Fig.1(b) the given unit-length, i.e., $z_0 = m\lambda$ where $\lambda$ is the wavelength labelled typically by wavenumber k and frequency $\omega$, its Hamiltonian can be written as follows,

$$H = \frac{1}{2}\int_0^{z_0}(L_l j^2(z) + C_l u^2(z))dz = 2L_l A^* A z_0$$

where

$$A = a\sqrt{\frac{\hbar\omega}{2L_l z_0}}, A^* = a^*\sqrt{\frac{\hbar\omega}{2L_l z_0}}.$$

According to the canonical quantization principle, we can quantize the system by operators $\hat{q}$ and $\hat{p}$, which satisfy the commutation relation $[\hat{q},\hat{p}]= i\hbar$. The annihilation and creation operators $\hat{a}$ and $\hat{a}^+$ are defined by the relations

$$\hat{a} = \frac{1}{\sqrt{2\hbar\omega}}(\omega\hat{q} + i\hat{p}), \hat{a}^\dagger = \frac{1}{\sqrt{2\hbar\omega}}(\omega\hat{q} - i\hat{p})$$

Thus the quantum Hamiltonian of Fig.1(b) can be rewritten as $\hat{H} = \hbar\omega\hat{a}^\dagger\hat{a} = \frac{1}{2}(\omega^2\hat{q}^2 + \hat{p}^2)$. Thus the current in the lossless unit cell equivalent circuit for LHTL can be quantized as

$$\hat{j}(z) = \sqrt{\frac{\hbar}{4\pi m L_l^2 C_l}}[\hat{a}\exp(\frac{i}{\omega\sqrt{C_l L_l}}z) + \hat{a}^\dagger\exp(-\frac{i}{\omega\sqrt{C_l L_l}}z)] \quad (4)$$

## 2 NRI in thermal FOCK state

The thermal noise would influence NRI when the LHTL operates at some temperature. In the following, we exploit NRI dependent thermal noise in thermal FOCK state. As for the equilibrium situation, the so-called thermo field dynamics (TFD) extends the usual quantum field theory to a finite temperature[36]. In TFD, the tilde space accompanies with the Hilbert space, and the tilde operators commute with the non-tilde operators[37]. Thus the creation and annihilation operators $\hat{a}^\dagger$, $\hat{a}$ associate with their tilde operators $\tilde{\hat{a}}^\dagger$, $\tilde{\hat{a}}$ according the rules[37]:

$$[\tilde{\hat{a}}, \tilde{\hat{a}}^\dagger] = 1, \tag{5}$$

$$[\tilde{\hat{a}}, \hat{a}] = [\tilde{\hat{a}}, \hat{a}^\dagger] = [\hat{a}, \tilde{\hat{a}}^\dagger] = 0 \tag{6}$$

The number operators in the Hilbert space and tilde space are read as $\hat{n} = \hat{a}^\dagger \hat{a}$, $\tilde{\hat{n}} = \tilde{\hat{a}}^\dagger \tilde{\hat{a}}$, In the direct product space, the thermal Fock state at finite temperature $|\hat{n}\tilde{\hat{n}}\rangle_T$ can be built by the thermal Bogoliubov transformation[37] through the Fock state at zero temperature $|\hat{n}\rangle \otimes |\tilde{\hat{n}}\rangle = |\hat{n}\tilde{\hat{n}}\rangle$ : $|\hat{n}\tilde{\hat{n}}\rangle_T = \hat{T}(\theta)|\hat{n}\tilde{\hat{n}}\rangle$, where $\hat{T}(\theta)$ is a thermal unitary operator which is defined as

$$\hat{T}(\theta) = exp[-\beta(\hat{a}\tilde{\hat{a}} - \hat{a}^\dagger \tilde{\hat{a}}^\dagger)] \tag{7}$$

the parameter $\theta$ is the thermal unitary operator relating the thermal photos $n_0$ in the thermal vacuum state: $n_0 = sinh\theta$. The thermal photos $n_0$ and temperature $T$ are ruled by the Boltzmann distribution $n_0 = [exp(\hbar\omega/k_B T) - 1]^{-1}$, in which $k_B$ is the Boltzmann constant. The bosonic operators in TFD can relate each other by the thermal Bogoliubov transformation as following,

$$\hat{T}^\dagger(\theta)\hat{a}\hat{T}(\theta) = \mu\hat{a} + \tau\tilde{\hat{a}}^\dagger, \tag{8}$$

$$\hat{T}^\dagger(\theta)\hat{a}^\dagger\hat{T}(\theta) = \mu\hat{a}^\dagger + \tau\tilde{\hat{a}} \tag{9}$$

where $\mu = cosh\theta$, $\tau = sinh\theta$. Then from Eq.(4) and Eq.(1), the quantum fluctuation of the current in the unit cell equivalent circuit is

$$\overline{(\Delta\hat{j})^2} = \frac{\hbar\omega^3\sqrt{\epsilon\mu}}{2z_0 Z_l}[2n_0^2 + 2(n+1)n_0 + 2n + 1] \tag{10}$$

where $n$ is the number of microwave field photons corresponding to the number operator $\hat{n}$ in the Hilbert space. The NRI according to the definition of the left-handed material $n_r = -\sqrt{\epsilon\mu}$ [1] can be written as follows,

$$n_r = -\frac{2z_0 Z_l \overline{(\Delta\hat{j})^2}[exp(\frac{\hbar\omega}{k_B T}) - 1][Cosh(\frac{\hbar\omega}{k_B T}) - 1]}{\hbar\omega^3\{[exp(\frac{\hbar\omega}{k_B T}) - 1][Cosh(\frac{\hbar\omega}{k_B T}) - 1] + 2n exp(\frac{\hbar\omega}{k_B T})[Cosh(\frac{\hbar\omega}{k_B T}) - 1] + exp(\frac{\hbar\omega}{k_B T}) - 1\}} \tag{11}$$

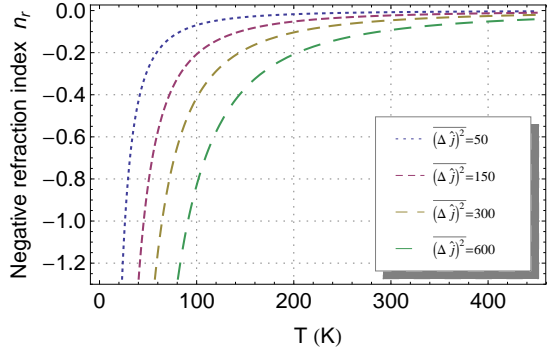

**Fig. 2** (Color online) Dependence of NRI $n_r$ and the temperature $T$ under the different current fluctuations in the LHTL.

## 3 Results and discussion

According Eq.(11), the expression of NRI is dependent temperature, frequency, photo numbers, and so on. However, the analytical expression of NRI in Eq.(11) corresponding to these parameters is rather cumbersome. Hence, we follow the numerical approach to get the dependence of NRI on these parameters.

Fig.2 shows NRI dependent the temperature $T$ under the different current fluctuations in one millimeter unit length of the LHTL with the characteristic impedance $Z_l = 50\Omega/m$, the photon numbers $n = 10$ and the frequency $\omega = 2GHz$. The destructive dependence of the temperature $T$ on NRI is shown when the current fluctuation varies from 50 to 600. And NRI declines sharply in the temperature interval of [0, 100K] when the current fluctuation $\overline{(\Delta\hat{j})^2}$=50. The phenomena demonstrates the ideal LHTL should operate at some low temperature $T$ to achieve a desired NRI. However, the quantum fluctuations of the current can improve the condition. The damping of NRI decreases with the increasing of $\overline{(\Delta\hat{j})^2}$ in the temperature interval of [0, 100K], and the damping (from -1.3 to -0.8) is minimum when the quantum fluctuation of the current $\overline{(\Delta\hat{j})^2}$=600 in the same temperature interval.

As mentioned before[22,34], NRI of the LHTL is achieved within the microwave frequency band. In the quantum mesoscopic LHTL, some novel feature of NRI is prominent within the microwave frequency band. Fig.3 shows NRI in the microwave frequency band [0, 3GHz] with $\overline{(\Delta\hat{j})^2}$=100 and photon numbers $n = 50$. It notes that NRI isn't homogeneous within the microwave frequency band, and NRI decreases sharply within [0, 1GHz] at an arbitrary temperature. A striking contrast between the different temperatures is obvious. The temperature of $T = 50K$ brings to the minimum damping of NRI within the frequency interval of [0, 1GHz], which demonstrates the lower frequency within the microwave frequency band is constructive to NRI, as coincides with the macroscopic LHTL[38].

As the energy of the electromagnetic wave is proportional to the photon numbers $n$ in the LHTL. In the mesoscopic quantizing LHTL, the influences of the electromagnetic energy on NRI is significant feature. Fig.4 illustrates NRI de-pendent photon numbers $n$ with $\overline{(\Delta \hat{j})^2}$=100. Different from Fig.2 and Fig.3, the temperature is much lower in Fig.4 than that in Fig.3. As it shown in Fig.4, the increasing photon numbers cause the decreasing NRI, and a larger NRI appears at a lower temperature. It also notes the values of NRI are much larger than those in Fig.2 and Fig.3 because of the lower temperature, which is shown by the ordinates in Fig.2, Fig.3 and Fig.4 describing the values of NRI.

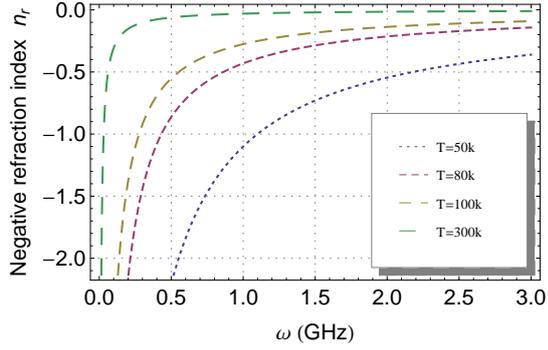

**Fig. 3** (Color online) Dependence of NRI $n_r$ and the frequency $\omega$ under different temperature in LHTL.

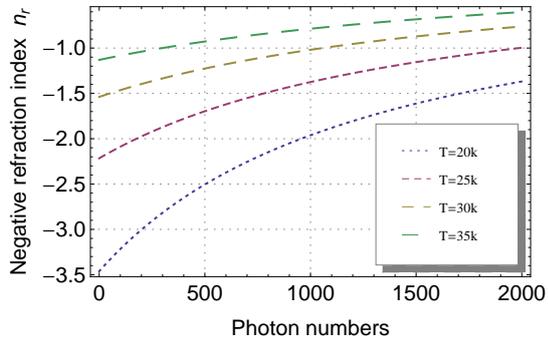

**Fig. 4** (Color online) Dependence of NRI $n_r$ and the photon numbers under different low temperatures in LHTL.

## 4 CONCLUSION

We discussed the quantum features of the NRI of the mesoscopic left-handed transmission line in the thermal Fock state. With the thermal fluctuation of current the NRI dependent the temperature, frequency of electromagnetic wave and photon numbers is discussed. The results show the lower temperature and frequency within the microwave frequency band, little photon numbers are more conducive to NRI, which is significant for the LHTL in miniaturizing applications in the coming future.

## References

1. V. Veselago, "The electrodynamics of substances with simultaneously negative values of $\varepsilon$ and $\mu$," *Soviet Physics Uspekhi* **10**, 509 (1968).
2. P. R. Berman, "Goos-Hänchen shift in negatively refractive media," *Phys. Rev. E* **66** 067603 (2002).
3. M. W. Feise, P. J. Bevelacqua and J. B. Schneider, "Effects of surface waves on the behavior of perfect lenses," *Phys. Rev.B* **66**, 035113 (2002).
4. K. Aydin, I. Bulu and E. Ozbay, "Subwavelength resolution with a negative-index metamaterial superlens," *Appl. Phys. Lett* **90**, 254102 (2007).
5. R. A. Shelby, D. R. Smith and S. Schultz, "Experimental verification of a negative index of refraction," *Science* **292**, 77 (2001)
6. H. J. Lezec, J. A. Dionne and H. A. Atwater, "Negative refraction at visible frequencies ," *Science* **316** 430 ( 2007 )
7. T. Xu, A. Agrawal, M. Abashin, K. J. Chau and H. J. Lezec, "All-angle negative refraction and active flat lensing of ultraviolet light," *Nature* **497** , 470 (2013)
8. H. Kosaka, T. Kawashima, A. Tomita, M. Notomi, T. Tamamura, T. Sato and S. Kawakami, "Superprism phenomena in photonic crystals",*Phys. Rev. B* **58** , R10096 (1998)
9. E. Cubukcu, K. Aydin, E. Ozbay, S. Foteinopoulou and C. M. Soukoulis, "Electromagnetic waves: Negative refraction by photonic crystals," *Nature* **423** , 604 (2003)
10. P. Parimi, W. Lu, P. Vodo, J. Sokoloff, J. Derov and S. Sridhar, "Negative refraction and left-handed electromagnetism in microwave photonic crystals," *Phys. Rev. Lett.* **92** , 127401 (2004)
11. J. B. Pendry, "A Chiral route to negative refraction," *Science* **306** 1353 (2004).
12. S. C. Zhao, Q. X. Wu, K. Ma, "Adjusting the left-handedness in a cold $^{87}$Rb atomic system via multiple parameter modulation," *Chinese Journal of Physics* **54**, 756 (2016).
13. Q. Thommen, P. Mandel, "Electromagnetically induced left handedness in optically excited four-level atomic media," *Phys. Rev. Lett.* **96**, 053601 (2006)
14. S. C. Zhao, S. Y. Zhang, Q. X. Wu, J. Jia, "Left-handedness with three zero-absorption windows tuned by the incoherent pumping field and inter-dot tunnelings in a GaAs/AlGaAs triple quantum dots system," *Chin. Phys. Lett.* **32(5)** 058104 (2015)
15. A. K. Iyer and G. V. Eleftheriades, "Negative refractive index metamaterials supporting 2-D waves," *IEEE-MTT Int'l Symp. 2 (412)*,(Seattle, WA, June 2002).
16. A. Grbic and G. V. Eleftheriades, "A backward-wave antenna based on negative refractive index L-C networks," *Proc. IEEE-AP-S USNC/URSI National Radio Science Meeting 4 (340)*, (San Antonio, TX, June 2002).
17. A. A. Oliner, "A periodic-structure negative-refractive-index medium without resonant elements," *URSI Digest, IEEE-AP-S USNC/URSI National Radio Science Meeting 41*, (San Antonio, TX, June 2002).
18. C. Caloz and T. Itoh, "Application of the transmission line theory of left-handed (LH) materials to the realization of a microstrip LH transmission line," *Proc. IEEE-AP-S USNC/URSI National Radio Science Meeting 2 (412)*, (San Antonio, TX, June 2002).
19. C. Caloz, H. Okabe, T. Iwai, and T. Itoh. "Anisotropic PBG surface and its transmission line model," *URSI Digest, IEEE-AP-S USNC/URSI National Radio Science Meeting,(224)*, (San Antonio, TX, June 2002).
20. R. E. Collin, *Foundations for Microwave Engineering(Second Edition)*, McGraw-Hill, (1992).
21. C. Caloz and T.Itoh, "Novel microwave devices and structures based on the transmission line approach of meta-materials," *IEEE-MTT Intl Symp. 1(195)*, Philadelphia, PA, June (2003).

22. Y. Horii, C. Caloz, and T. Itoh,"Super-compact multi-layered left-handed transmission line and diplexer application," *IEEE Trans. Microwave Theory Tech.*, **53**, 1527 (2005).
23. K. Abedi, "Improvement in performance of traveling wave electroabsorption modulator with asymmetric intra-step-barrier coupled double strained quantum wells the active region, segmented transmission-line and mushroom-type waveguide," *Opt Quant Electron* **44** 55 (2012)
24. A. Sanada, K. Murakami, I. Awai, H. Kubo, C. Caloz, and T. Itoh,"A planar zerothorder resonator antenna using a left-handed transmission line," *34th European Microwave Conference (1341)*,( Amsterdam, The Netherlands, Oct. 2004).
25. P. Y. G. Dontsop, B. G. O. Essama, J. M. Dongo et al. "Akhmediev-Peregrine rogue waves generation in a composite right/left-handed transmission line," *Opt Quant Electron* **48** 59 (2016).
26. A. Sanada, K. Murakami, S. Aso, H. Kubo, and I. Awai,"A via-free microstrip lefthanded transmission line," *IEEE-MTT Int'l Symp. (301)*, (Fort Worth, TX, June 2004).
27. S. Lim, C. Caloz and T. Itoh,"Metamaterial-based electronically-controlled transmission line structure as a novel leaky-wave antenna with tunable radiation angle and beamwidth," *IEEE Trans. Microwave Theory Tech.* **53** , 161 (2005).
28. N. A. Stathopoulos, S. P. Savaidis, and M. Rangoussi,"Propagation characteristics of nonlinear waveguides with complex refractive index using a transmission line model," *Opt Quant Electron* **38** 683 (2006)
29. F. A. Buot, "Mesoscopic physics and nanoelectronics: nanoscience and nanotechnology," *Phys. Rep.* **73**, 234 (1993).
30. D. J. Egger, F. K. Wilhelm, "Multimode circuit QED with hybrid metamaterial transmission lines," *Phys. Rev. Lett.* **111** , 163601 (2013).
31. R. G. Garcia, "Atomic-scale manipulation in air with the scanning tunneling microscope," *Appl. Phys. Lett.* **60** , 1960(1992).
32. Y. Makhlin, G. Schön, A. Shnirman, "JosephsonJunction qubits with controlled couplings," *Nature* **398** 305 (1999).
33. D. P. Divincenzo, D. Bacon, J. Kempe, G. Burkard, K.B. Whaley, "Universal quantum computation with the exchange interaction," *Nature* **408** 339 (2000).
34. C. Caloz and T. Itoh,"Transmission line approach of left-handed (LH) materials and microstrip implementation of an artificial LH transmission tine," *IEEE transactions. on antennas and propagation* **52** 1159 (2004).
35. W. H. Louisell, *Quantum statistical properties of radiation*, John Wiley, New York, (1973).
36. L. Laplae, F. Mancini and H. Umezawa, "Derivation and application of the boson method in superconductivity," *Physics Rep. C* **10** 151 (1974)
37. H. Umezawa, Y. Yamanaka,"Micro, macro and thermal concepts in quantum field theory," *Advances in Physics* **37** 531 (1988)
38. C. Caloz and T. Itoh, *Electromagnetic metamaterials: transmission line theory and microwave application*, John Wiley, New York, (2006).